# Poverty in the time of epidemic: A modelling perspective

Anand Sahasranaman[1,2,*] and Henrik Jeldtoft Jensen[2,3,#]

[1]Division of Sciences and Division of Social Sciences, Krea University, Sri City, AP 517646, India

[2]Centre for Complexity Science and Dept. of Mathematics, Imperial College London, London SW72AZ, UK.

[3]Institute of Innovative Research, Tokyo Institute of Technology, 4259, Nagatsuta-cho, Yokohama 226-8502, Japan.

[*] Corresponding Author. Email: anand.sahasranaman@krea.edu.in

[#] Email: h.jensen@imperial.ac.uk

**Abstract:**

We create a network model to study the spread of an epidemic through physically proximate and accidental daily human contacts in a city, and simulate outcomes for two kinds of agents - poor and non-poor. Under non-intervention, peak caseload is maximised, but no differences are observed in infection rates across poor and non-poor. Introducing interventions to control spread, peak caseloads are reduced, but both cumulative infection rates and current infection rates are systematically higher for the poor than for non-poor, across all scenarios. Larger populations, higher fractions of poor, and longer durations of intervention are found to progressively worsen outcomes for the poor; and these are of particular concern for economically vulnerable populations in cities of the developing world. Addressing these challenges requires a deeper, more rigorous understanding of the relationships between structural poverty and epidemy, as well as effective utilization of extant community level infrastructure for primary care in developing cities. Finally, improving iniquitous outcomes for the poor creates better outcomes for the whole population, including the non-poor.

# 1. Introduction:

In the wake of the novel coronavirus Covid-19 pandemic that is currently sweeping the planet, nations across the world have responded in many different ways. Non-pharmaceutical interventions (NPIs) to combat Covid-19 cover the entire spectrum of what is termed physical distancing – which is a set of methods to reduce frequency and closeness of contact between people to contain the spread of disease [1]. Some nations have issued advisories advocating that people maintain a minimum distance between each other as daily life goes on largely uninterrupted, while others have banned all large gatherings and enforced distancing norms more strictly, and yet others have even gone into 'lockdown' mode, requiring people to stay home and only allowing them to venture out to get basic supplies [2]. The spread of Covid-19 is being termed a pandemic, but most outbreaks are much more restricted to specific geographical locations, particularly in urban areas with greater contact density [3]. From the perspective of developing nations, therefore, attention needs to be focused on the impact of epidemics on those that are most vulnerable – the urban poor.

Prior work on understanding the link between poverty and epidemics indicates strong positive correlation between poverty and the fractions of population with communicable diseases across nations [4, 5]. While poverty is often considered to be a driver of disease, the nature of the relationship between poverty and communicable diseases does not suggest one-way causality, but rather a more complex relationship based on positive feedback [6, 7] . Poverty is found to create conditions for the spread of infectious diseases by forcing high density living and preventing access to medical infrastructure and care, which in turn means that diseases spread more easily in these communities and contributes to an exacerbation of poverty due to illness induced job losses, health costs, and mortality risks [6]. Infectious diseases have been found to have systematically affected economic development, with increasing burden of infectious diseases driven by falling biodiversity [7]. While these studies suggest that poverty and infectious disease reinforce each other, the relationship is largely explored at the level of nations. However, we are concerned about the spread of disease at finer-grained levels, specifically in cities of the developing world where a substantial proportion of the population is poor. In this context, the poor are characterized by a lack of sufficient economic resources that manifests in inability to meaningfully follow norms of distancing and contact reduction, or lack of access to the health services and pharmaceutical interventions; while the non-poor are able to follow norms and access health services.

In this work we attempt to model epidemic spread in an urban system (city) and estimate the differential impacts on poor and non-poor populations. We create a simple network model to propagate the stochastic dynamics of the spread of infection through a population consisting of two kinds of agents – poor and non-poor - and estimate the differential impacts on these populations. We explore model dynamics under various scenarios of intervention, and quantify the difference in outcomes across the two categories of agents during the course of the epidemic.

## 2. Model definition and specifications:

We model a network of a city consisting of $N$ agents, each a node in the network that represents the city. While the structure of social ties are well known to be approximated by power-law degree distributions [8], there remains considerable debate on the appropriate network structures to model the stochasticity inherent in the brief, physically proximate, and essentially accidental day-to-day encounters able to transmit an influenza-like infection [9, 10]. Given our interest in studying differential outcomes for poor and non-poor, and in the absence of data on networks of spread in developing cities, we propose to explore the dynamics of transmission on two different idealized network structures – an uncorrelated random Erdős–Rényi network [11] and a scale-free Barabási-Albert network [12].

In the Erdős–Rényi (ER) graph, each edge is generated at random with probability $p_{link}$, and is independent of all other edges. Essentially, a node's neighbours on the network represent their daily contacts - those they come into close contact physically (at home, at public conveniences, in transit, at work etc.) during the course of a day - and given the inherent randomness in many of these proximate contacts, the ER network is arguably a reasonable representation of this reality. A given value of $p_{link}$ yields an average of $q = (N-1)p_{link}$ neighbours for a random node in the network, resulting in an approximately Poisson degree distribution (as $N \to \infty$ and $Np_{link}$ is constant) that is indicative of the lack of clustering in the network and homogeneity of node-level network properties.

For the Barabási-Albert (BA) network, a graph with $N$ nodes is obtained by attaching new nodes with $m$ neighbours, based on preferential attachment. The degree distribution for such a network follows a power-law, meaning that the average degree of a node in the network hides significant heterogeneity – many nodes have a smaller number of connections, but a few have much larger numbers of connections. Such highly connected super-spreader nodes are now considered critically important in the context of the dynamics of disease transmission [13, 14].

At the outset, when time $t = 0$ days, each node is randomly designated as being poor or non-poor for the duration of the dynamics based on Eq. 1:

$$S(n_i) = \begin{cases} 0, & w.p. \ p_{poor} \\ 1, & w.p. \ 1 - p_{poor} \end{cases}, \tag{1}$$

where $S(n_i)$ is the status of node $i$. Each time increment in the dynamics represents a single day.

We model the epidemic using the classic Kermack-McKendrick three-compartment Susceptible-Infected-Recovered (SIR) model [15]. The SIR model and its variants are commonly used to understand the progression of an epidemic and its impact on the health system. We use this construct instead to focus attention on the differential nature of impact of the epidemic on underlying populations of the poor and non-poor.

When $t = 0$, there is one randomly infected node and $N - 1$ susceptible nodes. Given that the transmission probability from an infected to a susceptible node is $p$, at each iteration $t$, an infected node with contact rate $k$ causes, on average, $\beta$ infections. $\beta$ is the daily transmission rate and is given by (Eq. 2):

$$\beta = pk \tag{2}$$

Once the transmission dynamics for an iteration are completed, each infected agent that has spent $1/\gamma$ days being infected, moves into the recovered compartment. A recovered agent is inert, in that it is neither susceptible nor infective. Model dynamics are propagated a period of $t = T_f$ iterations (days).

The base reproductive number $R_0$ is a measure of the number of infections caused by an infected individual, given that the population has no immunity from past exposures or vaccination, nor any deliberate intervention in disease transmission [16] and is given by (Eq. 3):

$$R_0 = \beta/\gamma, \tag{3}$$

The effective reproduction rate, $R_e(t)$, at any time $t$ during the spread of infection depends on the contact rates and transmission probabilities at that time, and these variables co-evolve with the numbers of susceptible, infected, and recovered individuals in the system. This co-evolution is impacted by the nature and extent of both pharmaceutical and non-pharmaceutical interventions [17] implemented to contain spread of disease. We study the evolution of outcomes for the poor and non-poor under three scenarios of intervention:

  i. *Non-intervention:* a scenario where we assume that no measures are taken to combat transmission and contact, and the epidemic infects the population and runs the entire course of dynamics with constant transmission probability $p$ and an unchanged network of connections.
  ii. *Transmission rate reduction:* this intervention refers to strategies such as pharmaceutical interventions including medicines as well as non-pharmaceutical norms such as physical distancing or masking to reduce transmission. However, given the constraints of the poor in accessing or implementing these strategies, we see differential transmission probabilities for the poor and non-poor. Essentially, the transmission probability from non-poor nodes to all other nodes (poor and non-poor) reduces to $p_{np-np} = p_{np-p} = p_{np}$ ($p_{np} < p$), indicating their ability to access or follow the intervention strategy. Meanwhile, transmission probability between poor and non-poor nodes also decreases to $p_{p-np} = p_{np}$ ($p_{np} < p$), because the non-poor are still able to implement the intervention and protect themselves, but the transmission probability from poor nodes to other poor nodes is $p_{p-p} = p_p$ ($p_{np} < p_p \leq p$), reflecting the difficulties and constraints faced by the poor in accessing requisite medication or medical care. Overall, while effective transmission rates of both the non-poor and poor nodes decrease under this intervention, the effect is smaller for the poor nodes vis-à-vis non-poor nodes.

iii. *Contact rate reduction:* this scenario refers to stronger interventions such as lockdown or highly stringent distancing norms, that alter the network of daily proximate contacts of individuals by reducing connectivity in the network, and thus containing spread. Again, the poor are not able to reduce contacts as effectively as the non-poor because of structural conditions such as living conditions as well as being employed in essential sectors such as health and sanitation. We model the differential impact on the poor and non-poor by reducing the average number of contacts from a non-poor node to any other (poor or non-poor) node to a proportion: $q_{np-p} = q_{np-np} = q_{np}$ ($q_{np} < 1$) of its neighbours prior to intervention. The contact rate from a poor node to a poor node is reduced to a proportion $q_{p-p} = q_p$ ($q_{np} < q_p \leq 1$) of its neighbours prior to intervention, indicating the greater difficulty of poor agents in reducing contacts. The contact rate of a poor node to a non-poor node is also reduced to $q_{p-np} = q_{np}$, on account of the non-poor nodes reducing their contacts with the poor.

We test outcomes across all these scenarios for both network types under consideration. Table 1 provides the complete set of parameter values and initial conditions for the simulations.

|  | Values (Erdős–Rényi) | Values (Barabási-Albert) |
|---|---|---|
| **Parameters** | | |
| Population – number of network nodes, $N$ | 10,000 | 10,000 |
| Edge probability, $p_{link}$ (Erdős–Rényi) | 0.005 | |
| Number of edges from new node to extant nodes, $m$ (Barabási-Albert) | | 50 |
| Probability of agent being poor, $p_{poor}$ | 0.50 | 0.50 |
| Transmission probability, $p$ | 0.005 | 0.005 |
| Recovery rate, $\gamma$ | 0.10 | 0.10 |
| Number of iterations (days) in one simulation of model, $T_f$ | 200 | 200 |
| Number of simulations | 100 | 100 |

| Scenarios | | |
|---|---|---|
| Transmission probability from poor to poor under intervention II, $p_{p-p}$ | 0.005 | 0.005 |
| Transmission probability from poor to non-poor under intervention II, $p_{p-np}$ | 0.003 | 0.003 |
| Transmission probability from non-poor (to both poor and non-poor) under intervention II, $p_{np}$ | 0.003 | 0.003 |
| Contact rate for poor to poor under intervention III, $q_{p-p}$ | 0.8 | 0.8 |
| Contact rate for poor to non-poor under intervention III, $q_{p-np}$ | 0.4 | 0.4 |
| Contact rate for non-poor (with both poor and non-poor) under intervention III, $q_{np}$ | 0.4 | 0.4 |
| Start of intervention ($t_{int}$) | 40 | 15 |
| Duration of intervention ($t_{dur}$) | 60 | 30 |
| **Initial Conditions** | | |
| Number of susceptible nodes, $S(0)$ | 9,999 | 9,999 |
| Number of infectious nodes, $I(0)$ | 1 | 1 |
| Number of recovered nodes, $R(0)$ | 0 | 0 |

**Table 1**: *Parameter values and initial conditions.*

The choice of these base parameter values (Table 1) for network structure, $p$, and $\gamma$ yields $R_0 = 2.5$. This regime ($R_0 > 1$) simulates an epidemic propagating through the network, enabling us to explore dynamics of spread. However, we vary these assumptions in alternative simulations of the model by changing $\gamma$ ($0.05 \leq \gamma \leq 0.2$), thus yielding $1.25 \leq R_0 \leq 5$, and enabling assessment of model sensitivity to $R_0$. We also vary the system size across 3 orders of magnitude ($1000 \leq N \leq 100,000$) and the proportion of poor significantly ($0.25 \leq p_{poor} \leq 0.75$) to estimate the robustness of our findings. Finally, we test the model by varying the start time and duration of intervention. Outcomes of these various alternative parametrizations of the model are discussed in Section 5.

Fig. 1 depicts representative structures of the ER and BA graphs used in this model, portraying scaled-down networks of connections with 100 nodes (instead of 10,000) for the sake of visual clarity. As is apparent, there is much lesser variance in the degree distribution of nodes in the ER (Fig. 1a), while the BA graph has a dense core (darker shade of edges representing greater connectedness) containing a small set of nodes with much higher degree than the average degree for nodes in the network (Fig. 1b). These networks represent an equal number of poor (yellow nodes) and non-poor (purple nodes), with no systematic bias in degree distribution based on poverty status.

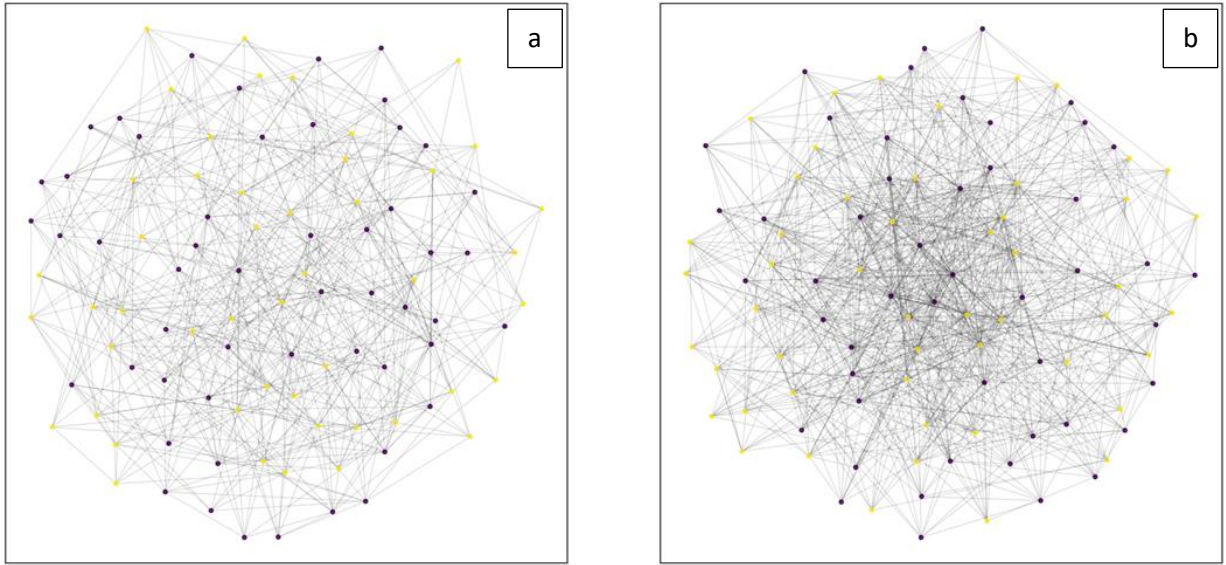

**Figure 1:** *Representative structures of ER and BA graphs*. a: Erdős–Rényi (ER) graph with 100 nodes and link probability of 0.1. b. Barabási-Albert (BA) graph with 100 nodes and 10 edges attached from each new node to existing nodes. The purple nodes represent non-poor agents and the yellow nodes represent poor agents.

We are interested in two sets of outcomes – one is the current infectious caseload at any given time $t$, and the other is the cumulative infections over time. We measure both these outcomes not only for the overall population, but also the poor and non-poor populations distinctly so as to assess the differential impact of the epidemic and interventions applied. If, at a given time $t$, $f_I(t)$, $f_I(p,t)$ and $f_I(np,t)$ are the fractions of overall, poor and non-poor populations that have ever been infected, then Differential Impact ($DI$) is defined as the ratio of the cumulative fraction of poor infected to non-poor infected (Eq. 4).

$$DI = f_I(p, T_f)/f_I(np, T_f) \qquad (4)$$

## 3. Analytical description of dynamics:

Given the model in Section 2, we present an analytical treatment of the dynamics. It is important to highlight at the outset that this analytical mean-field description is only suited to approximate the dynamical outcomes from the ER graph because of its underlying homogeneity, and not for the heterogenous network structure obtained in the BA model.

The total population of the system is $N$, and based on the edge probability $p_{link}$, each node has, on average, $q = (N-1)p_{link}$ neighbours; this is the average contact rate of a node in this network. The probability of transmission of disease from an infected individual to a susceptible individual is $p$; therefore, the average number of infections caused by an individual per day is $\beta = pq$. Also, given $p_{poor}$, we have, on average, a fraction $f_p$ of the population that is poor and a corresponding fraction $f_{np} = 1 -$

$f_p$ that is non-poor. At the end of a time interval $t$, fractions $f_S(t)$ of the overall population, $f_S(p,t)$ of the poor population, and $f_S(np,t)$ of the non-poor population, are still susceptible ($S$).

In the *non-intervention* scenario, where transmission in a random network is uncontained, on average, $f_S(p,t) = f_S(np,t) = f_S(t)$. The effective transmission rate from an infected individual at time $t$, $R_e(t)$, taking into account the current state of the network is given by (Eq. 5):

$$R_e(t) = p f_S(t-1) q / \gamma \tag{5}$$

Using Eq. 2, we have (Eq. 6):

$$R_e(t) = \beta f_S(t-1)/\gamma \tag{6}$$

At $t=0$ and in the limit $N \to \infty$, $R_e(0) = R_0 = \beta/\gamma$. The susceptible fraction of the population is (Eq. 7):

$$f_S(t) = 1 - f_I(t), \tag{7}$$

where $f_I(t)$ is the fraction of the overall population ($N$) that has ever been infected, and is given by (Eq. 8):

$$f_I(t) = \begin{cases} \frac{1}{N}, & \text{for } t = 0 \\ f_I(t-1) + (\beta f_I(t-1)(1 - f_I(t-1)), & \text{for } 1 \leq t < 1/\gamma \\ f_I(t-1) + (\beta \left(f_I(t-1) - f_I\left(t - \frac{1}{\gamma}\right)\right)(1 - f_I(t-1)), & \text{for } t \geq 1/\gamma \end{cases} \tag{8}$$

Given that $f_S(p,t) = f_S(np,t)$ in unconstrained transmission, the effective transmission rate for poor nodes $R_e(p,t)$ and non-poor nodes $R_e(np,t)$ are, on average, equal (Eq. 9):

$$R_e(p,t) = R_e(np,t) = R_e(t) = \beta f_S(t-1)/\gamma \tag{9}$$

Next, we consider the scenario of *transmission rate reduction*. Under this intervention, we need to compute appropriate $\beta$ for different kinds of connections - non-poor to non-poor ($np \to np$), non-poor to poor ($np \to p$), poor to non-poor ($p \to np$), and poor to poor ($p \to p$). Here, $\beta$ changes in accordance with changes in $p$, keeping $q$ constant (Eq. 10).

$$\beta = \begin{cases} \beta_{np-np} = p_{np-np}q = p_{np}q = \beta_{np}, & \text{if } np \to np \\ \beta_{np-p} = p_{np-p}q = p_{np}q = \beta_{np}, & \text{if } np \to p \\ \beta_{p-np} = p_{p-np}q = p_{np}q = \beta_{np}, & \text{if } p \to np \\ \beta_{p-p} = p_{p-p}q = p_{p}q = \beta_{p}, & \text{if } p \to p \end{cases}$$

Given $f_p$ and $f_{np}$, we can estimate the transmission probability (Eq. 11).

$$p = f_{np}^2 \frac{\beta_{np}}{q} + f_{np}f_p \frac{\beta_{np}}{q} + f_p f_{np} \frac{\beta_{np}}{q} + f_p^2 \frac{\beta_p}{q} = (f_{np}^2 + 2f_p f_{np}) \frac{\beta_{np}}{q} + f_p^2 \frac{\beta_p}{q} \tag{11}$$

And the effective transmission rate $R_e(t)$ is (Eq. 12):

$$R_e(t) = \left((f_{np}^2 + 2f_p f_{np})\beta_{np} + f_p^2 \beta_p\right)\left(f_{np}f_S(np, t-1) + f_p f_S(p, t-1)\right)/\gamma \tag{12}$$

$R_e(t)$ estimates the overall effective transmission rate, but we would also like to understand effective transmission rates of poor nodes $R_e(p, t)$ and non-poor nodes $R_e(np, t)$ respectively (Eqs. 13, 14).

$$R_e(p, t) = \left(\beta_{np}f_{np}f_S(np, t-1) + \beta_p f_p f_S(p, t-1)\right)/\gamma \tag{13}$$

$$R_e(np, t) = \beta_{np}\left(f_{np}f_S(np, t-1) + f_p f_S(p, t-1)\right)/\gamma \tag{14}$$

In this case, fraction of susceptible nodes are given as (Eq. 15):

$$f_S(t) = 1 - f_I(t); f_S(p, t) = 1 - f_I(p, t); f_S(np, t) = 1 - f_I(np, t) \tag{15}$$

And the infected node fraction is estimated by (Eq. 16):

$$f_I(t) = f_p f_I(p, t) + f_{np} f_I(np, t) \tag{16}$$

The fraction of poor and non-poor infected nodes, on average, are given as (Eqs. 17, 18):

$$f_I(p, t) = \begin{cases} \frac{1}{N}, & \text{for } t = 0 \\ f_I(p, t-1) + \beta f_p f_I(p, t-1)(1 - f_I(p, t-1)) + \beta f_{np} f_I(np, t-1)(1 - f_I(np, t-1)), & \text{for } 1 \le t < \frac{1}{\gamma} \\ f_I(p, t-1) + (\beta f_p \left(f_I(p, t-1) - f_I\left(p, t - \frac{1}{\gamma}\right)\right)(1 - f_I(p, t-1)) + \beta f_{np} \left(f_I(np, t-1) - f_I\left(np, t - \frac{1}{\gamma}\right)\right)(1 - f_I(np, t-1)), & \\ & \text{for } 1/\gamma \le t \le t_{int} - 1 \text{ and } t \ge t_{int} + t_{dur} \\ f_I(p, t-1) + (\beta_p f_p \left(f_I(p, t-1) - f_I\left(p, t - \frac{1}{\gamma}\right)\right)(1 - f_I(p, t-1)) + \beta_{np} f_{np} \left(f_I(np, t-1) - f_I\left(np, t - \frac{1}{\gamma}\right)\right)(1 - f_I(np, t-1)), & \\ & \text{for } t_{int} \le t < t_{int} + t_{dur} \end{cases} \tag{17}$$

$$f_I(np, t) = \begin{cases} \frac{1}{N}, & \text{for } t = 0 \\ f_I(np, t-1) + \beta f_p f_I(p, t-1)(1 - f_I(p, t-1)) + \beta f_{np} f_I(np, t-1)(1 - f_I(np, t-1)), & \text{for } 1 \le t < 1/\gamma \\ f_I(np, t-1) + (\beta f_p \left(f_I(p, t-1) - f_I\left(p, t - \frac{1}{\gamma}\right)\right)(1 - f_I(p, t-1)) + \beta f_{np} \left(f_I(np, t-1) - f_I\left(np, t - \frac{1}{\gamma}\right)\right)(1 - f_I(np, t-1)), & \\ & \text{for } 1/\gamma \le t \le t_{int} - 1 \text{ and } t \ge t_{int} + t_{dur} \\ f_I(np, t-1) + (\beta_{np} f_p \left(f_I(p, t-1) - f_I\left(p, t - \frac{1}{\gamma}\right)\right)(1 - f_I(p, t-1)) + \beta_{np} f_{np} \left(f_I(np, t-1) - f_I\left(np, t - \frac{1}{\gamma}\right)\right)(1 - f_I(np, t-1)), & \\ & \text{for } t_{int} \le t < t_{int} + t_{dur} \end{cases} \tag{18}$$

where $t = t_{int}$ indicates the time when the intervention begins and $t = t_{int} + t_{dur}$ is the time when the intervention ends.

Finally, we consider the *contact rate reduction scenario*, where the number of contacts of an agent is differentially reduced across rich and poor. Here, $\beta$ changes in accordance with changes in $q$, keeping transmission probability $p$ constant (Eq. 19):

$$\beta = \begin{cases} \beta_{np-np} = pq_{np-np} = pq_{np} = \beta_{npq}, & if\ np \to np \\ \beta_{np-p} = pq_{np-p} = pq_{np} = \beta_{npq}, & if\ np \to p \\ \beta_{p-np} = pq_{p-np} = pq_{np} = \beta_{npq}, & if\ p \to np \\ \beta_{p-p} = pq_{p-p} = pq_p = \beta_{pq}, & if\ p \to p \end{cases} \quad (19)$$

The effective transmission rate under this scenario is (Eq. 20):

$$R_e(t) = \left(\left(f_{np}^2 + 2f_p f_{np}\right)\beta_{npq} + f_p^2 \beta_{pq}\right)\left(f_{np} f_S(np, t-1) + f_p f_S(p, t-1)\right)/\gamma \quad (20)$$

We estimate the effective transmission rates of poor and non-poor nodes (Eqs. 21, 22):

$$R_e(p, t) = (\beta_{pq} f_p f_S(p, t-1) + \beta_{npq} f_{np} f_S(np, t-1))/\gamma \quad (21)$$

$$R_e(np, t) = \beta_{npq}(f_p f_S(p, t-1) + f_{np} f_S(np, t-1))/\gamma \quad (22)$$

We use Eq. 15 to estimate the susceptible fractions of population, based on the infected fractions and Eq. 16 to estimate the infected fraction of overall population. Infected fractions of poor and non-poor categories are given by (Eqs. 23, 24):

$f_I(p, t) =$

$$\begin{cases} \frac{1}{N}, & for\ t = 0 \\ f_I(p, t-1) + \beta f_p f_I(p, t-1)(1 - f_I(p, t-1)) + \beta f_{np} f_I(np, t-1)(1 - f_I(np, t-1)), & for\ 1 \leq t < \frac{1}{\gamma} \\ f_I(p, t-1) + (\beta f_p \left(f_I(p, t-1) - f_I\left(p, t-\frac{1}{\gamma}\right)\right)(1 - f_I(p, t-1)) + \beta f_{np} \left(f_I(np, t-1) - f_I\left(np, t-\frac{1}{\gamma}\right)\right)(1 - f_I(np, t-1)), & \\ & for\ 1/\gamma \leq t \leq t_{int} - 1\ and\ t \geq t_{int} + t_{dur} \\ f_I(p, t-1) + (\beta_{pq} f_p \left(f_I(p, t-1) - f_I\left(p, t-\frac{1}{\gamma}\right)\right)(1 - f_I(p, t-1)) + \beta_{npq} f_{np} \left(f_I(np, t-1) - f_I\left(np, t-\frac{1}{\gamma}\right)\right)(1 - f_I(np, t-1)), & \\ & for\ t_{int} \leq t < t_{int} + t_{dur} \end{cases} \quad (23)$$

$f_I(np, t) =$

$$\begin{cases} \frac{1}{N}, & for\ t = 0 \\ f_I(np, t-1) + \beta f_p f_I(p, t-1)(1 - f_I(p, t-1)) + \beta f_{np} f_I(np, t-1)(1 - f_I(np, t-1)), & for\ 1 \leq t < 1/\gamma \\ f_I(np, t-1) + \beta f_p \left(f_I(p, t-1) - f_I\left(p, t-\frac{1}{\gamma}\right)\right)(1 - f_I(p, t-1)) + \beta f_{np} \left(f_I(np, t-1) - f_I\left(np, t-\frac{1}{\gamma}\right)\right)(1 - f_I(np, t-1)), & \\ & for\ 1/\gamma \leq t \leq t_{int} - 1\ and\ t \geq t_{int} + t_{dur} \\ f_I(np, t-1) + \beta_{npq} f_p \left(f_I(p, t-1) - f_I\left(p, t-\frac{1}{\gamma}\right)\right)(1 - f_I(p, t-1)) + \beta_{npq} f_{np} \left(f_I(np, t-1) - f_I\left(np, t-\frac{1}{\gamma}\right)\right)(1 - f_I(np, t-1))), & \\ & for\ t_{int} \leq t < t_{int} + t_{dur} \end{cases} \quad (24)$$

## 4. Simulated results and dynamics:

When we study the current infected caseload over time on the ER graph in the *non-intervention scenario*, we find the classic exponential curve described the by the dynamics in Fig. 2a, as expected in the SIR model [15]. Essentially, this means that there is an exponential increase in infected caseloads, with the potential to significantly impact and even overwhelm the health system. We also explore the difference in proportions of poor and non-poor agents (in terms of their respective populations) that form this caseload, and find that both sections of the population are similarly impacted (Fig. 2a). This is brought into sharper relief when we study the cumulative proportions of poor and non-poor affected over time and find that the temporal progression of the proportions of both populations follow exactly the same trajectory (Fig. 2b). At the end of the outbreak ~82% of the overall population has been infected, with the exact same proportion reflected in both poor and non-poor categories. The analytical description of the model indicates reasonable correspondence with these simulation results (Eqs. 6-8), estimating that ~87% of the population is infected, with the same levels of infection across both categories. Given that we have a random network and unconstrained spread, this result is in line with expectations. Qualitatively similar behaviour is observed with the exponential rise of infections in the BA network leaving both poor and non-poor equally affected (Figs. 2c and 2d) - though the overall outcomes are much worse than on the ER graph, both in terms of peak caseloads and cumulative population infected. The peak of the epidemic yields significantly higher infections at ~69%, and while this means that the epidemic runs its course much quicker, it also affects ~96% of the overall population. This difference in outcomes compared to the ER network is attributable to the role of nodes with higher connections being infected and, in turn, spreading the infection widely in the network.

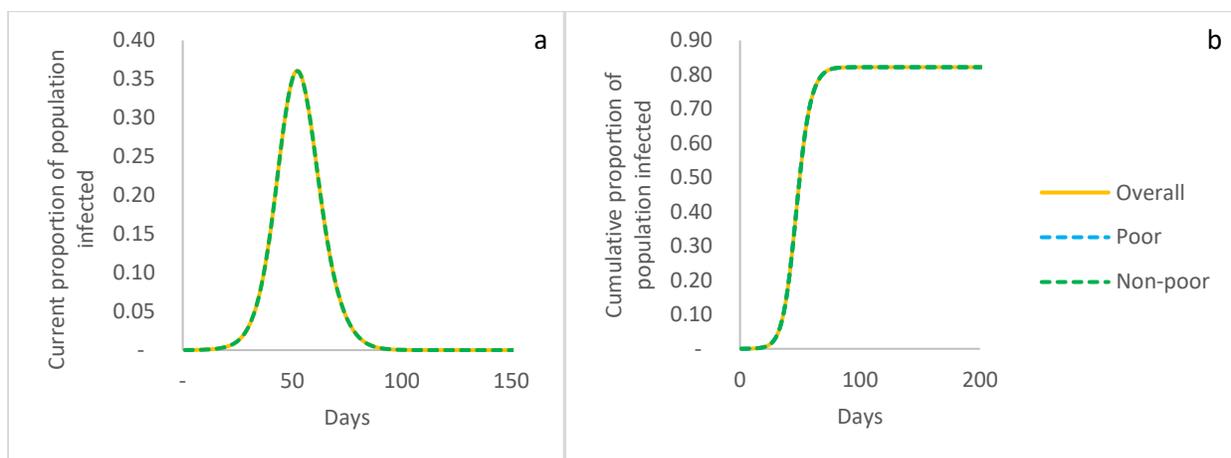

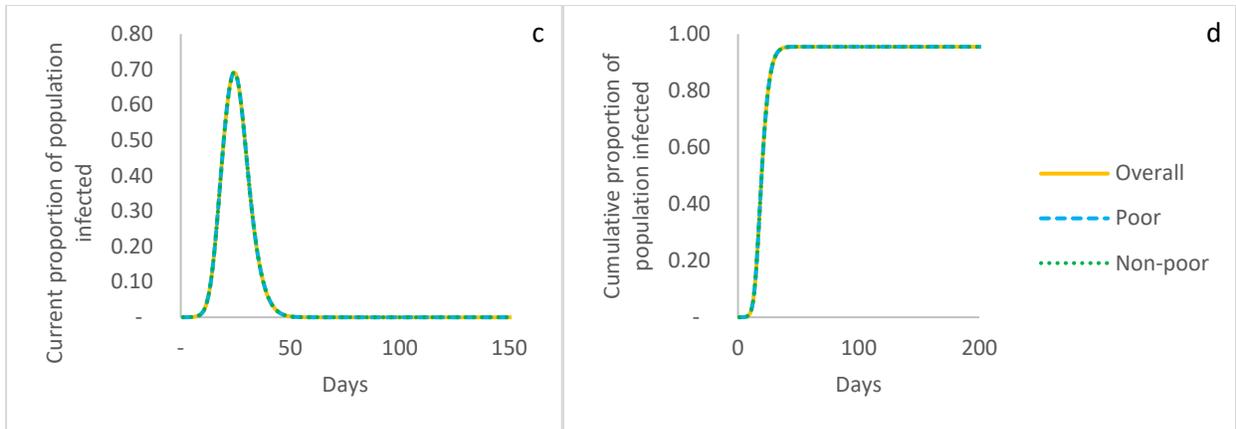

**Figure 2:** *Simulated system outcomes in non-intervention scenario*. Yellow line represents overall population; dashed blue line represents poor population; and dashed green line represents non-poor population. a: Current proportion of population infected over time on ER graph. Depicts the current infected caseload (as of each day) as a proportion of the population. All sections of the population contribute proportionally to caseloads. b: Cumulative proportion of population infected over time in ER graph. Depicts the temporal evolution of cumulative populations infected. All sections of the population display the same trend in cumulative infection rates. c: Current proportion of population infected over time in BA graph. Shows exponential growth and decay with a much higher peak than in the ER graph. b: Cumulative proportion of population infected over time in BA graph. Fractions of population affected are exactly the same across poor and non-poor categories, though much higher than in ER graph. Overall, in this non-intervention scenario, both categories of the population are similarly affected, per expectation, across all network types.

We now turn our attention to the *transmission probability reduction scenario*. In terms of overall trends, we find, as expected, that the peak infection rate (25%) and cumulative population infected (73%) in the ER graph under this scenario are lower than in the non-intervention scenario (36% and 82% respectively), because at time $t = 41$, the intervention that reduces the effective transmission rate begin to take effect (Figs. 3a and 3b). Similar trends are apparent in the BA network as well with reduced peak infection rate (55%) and cumulative infections (87%) under this intervention when compared to non-intervention (69% and 96% respectively) (Figs. 3c and 3d). However, we now begin to observe the differential impacts of the intervention on poor and non-poor populations, which show distinctive paths post $t = 40$, when intervention is initiated. In the ER graph, we see that the poor population shows a higher infected caseload peak at 27% occurring at $t = 49$ (against 23% for the non-poor population occurring at $t = 47$), as a result of having a higher effective transmission rate post intervention. Consequently, we also see that the cumulative population infected in the case of the poor is higher at 76%, as against 69% of non-poor (Fig. 3b), making clear the unequal impacts of the intervention on different sections of the city's population. Correspondingly, in the BA graph, we see the peak caseload for the non-poor population to be 53% as against 58% for the poor population, and cumulative proportions of populations infected to be 89% and 86% respectively (Figs. 3c and 3d).

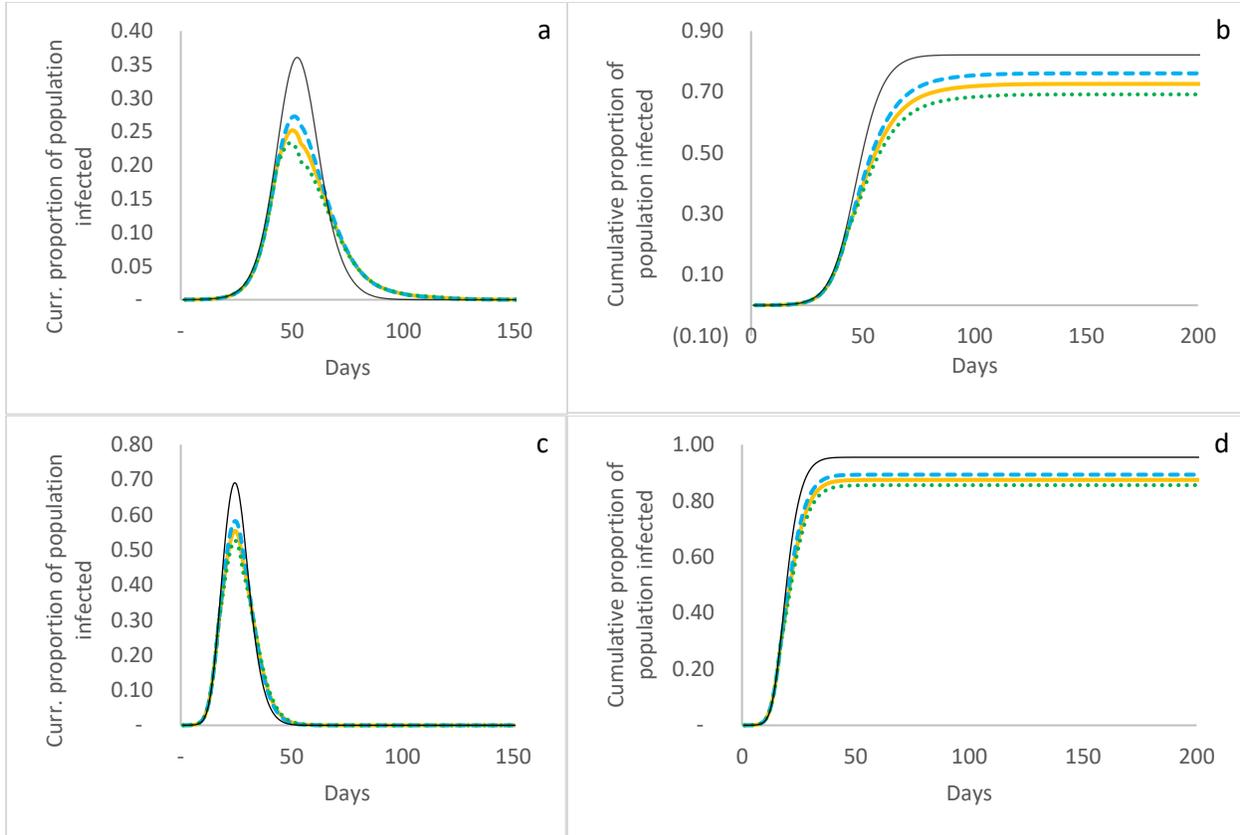

**Figure 3:** *Simulated system outcomes in transmission probability reduction scenario.* Yellow line represents overall population; dashed blue line represents poor population; dashed green line represents non-poor population; and thin black line represents non-intervention scenario for comparison. a: Current proportion of population infected over time on ER graph. Peak overall caseload is lower than non-intervention scenario, but poor population has a higher peak than non-poor. b: Cumulative proportion of population infected over time on ER graph. Cumulative infected is lower than under non-intervention, but the poor suffer a higher cumulative infection rate than the non-poor (76% v. 69%). c: Current proportion of population infected over time on BA graph. As in the case of the ER graph outcomes, peak caseload is lower than non-intervention scenario, but with a higher peak for the poor. b: Cumulative proportion of population infected over time on BA graph. Overall infected is lower than under non-intervention, with poor having an overall higher cumulative infection rate (89% v. 86%).

We now explore the temporal evolution of the effective transmission rate $R_e(t)$ based on the analytical description of the model, and compare it to simulated outcomes on the ER graph (Fig. 4). As per Eq. 12, the effective transmission rate at $t = 41$ would be $R_e(t = 41) = 1.4$, as against a value of $R_e(t = 41) = 2.0$ in the case of non-intervention (Eq. 6). This means that disease transmission becomes slower and there is a lower peak at $t = 48$, compared to the later and higher peak at $t = 51$ for the non-intervention case (Fig 2a), but that now the overall duration of the infection is more spread out. We find that our mean field estimates (Eqs. 12-14) of the point in time when effective transmission rates for all population categories start declining just below 1 are in broad concurrence with attainment of peak caseloads observed in simulations – they occur at time period $t \sim 50$ for overall population (48 in simulations), at $t \sim 46$ for non-poor (47 in simulations), and at $t \sim 53$ for poor (49 in simulations) (Fig. 4). The analytical model also estimates that after the end of intervention at $t_{int} + t_{dur}$, the effective transmission rate shows an increase (but is still below 1) because transmission probability and contact rates are now back to pre-intervention levels, but given the state of the system in terms of numbers of recovered agents (who are now immune) and the remaining susceptible population, there is no further increase in infections.

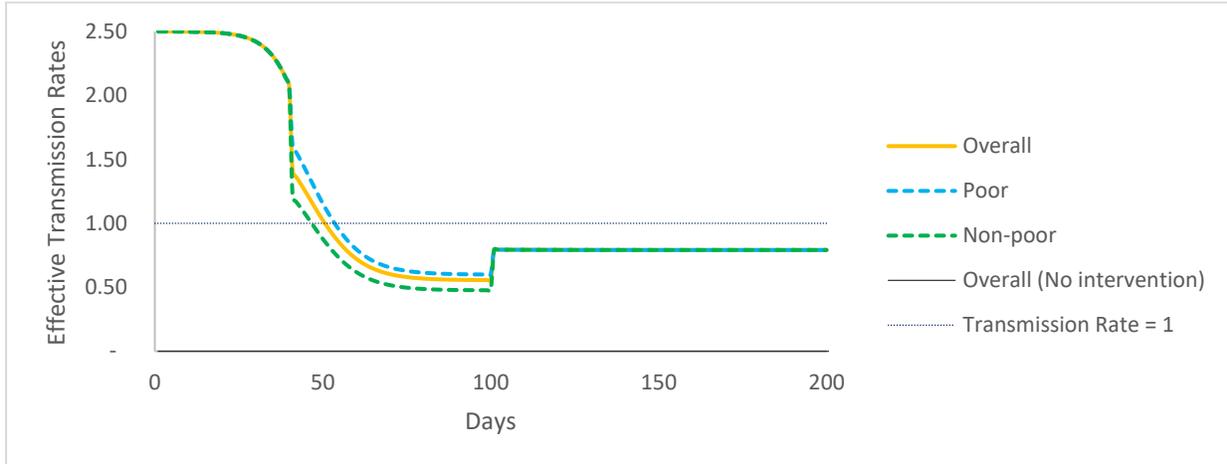

**Figure 4:** *Evolution of mean-field effective transmission rate under transmission rate reduction scenario:* Until the intervention begins, effective transmission rate of all sections of the population follow the same trajectory, but then show differing paths with transmission rates for poor higher during the period of intervention. Post-intervention, again, the trajectories merge and effective transmission rate remains flat, under 1.

Finally, we consider the outcomes under the *contact rate reduction scenario*. On the ER network, we find that peak caseload is significantly lower than the non-intervention scenario (21% v. 36%), but just like in the case of the transmission probability reduction scenario, the poor have a higher peak than the non-poor (23% v. 20%) (Fig. 5a). However, there is a significantly larger difference in the cumulative fraction of population infected – it is 61% for the non-poor against 68% for the poor (Fig. 5b). On the BA network, we find that the peak caseload for the poor is 53% as against 46% for the non-poor (Fig. 5c), and the cumulative fraction infected is also significantly higher for the poor than the non-poor (88% and 81%) (Fig. 5d).

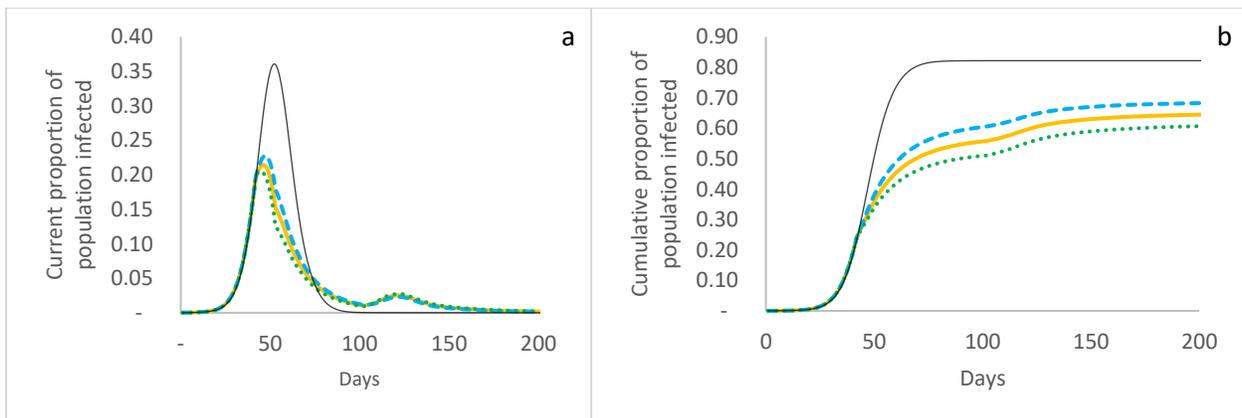

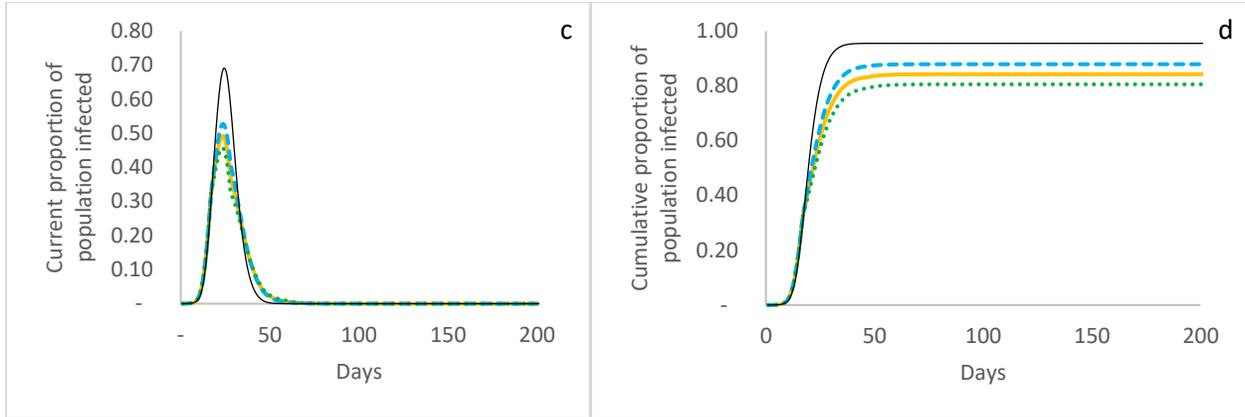

**Figure 5:** *Simulated system outcomes for contact rate reduction scenario*. Yellow line represents overall population; dashed blue line represents poor population; dashed green line represents non-poor population; and thin black line represents non-intervention scenario for comparison. a: Current proportion of population infected over time on ER graph. Peak overall caseload is lower than non-intervention scenario with poor population having a higher peak than non-poor, but we see a second peak post intervention, where non-poor are also affected. b: Cumulative proportion of population infected over time on ER graph. Cumulative infected is lower than under non-intervention, but the poor suffer a much higher cumulative infection rate than the non-poor (68% v. 61%). c: Current proportion of population infected over time on BA graph. Poor have higher peak than non-poor, but overall peak is lower than under non-intervention. d: Cumulative proportion of population infected over time on BA graph. Poor have much higher cumulative infection rate than the non-poor (88% v. 81%).

An interesting aspect revealed in the dynamics on the ER graph is that once the intervention ends after $t_{dur}$ days, the fraction of current infections starts rising once again resulting in a smaller second peak of infections, which affects the non-poor worse than the poor (Fig. 5a). This is because at the end of the intervention, there is a non-zero proportion of the population that remains infected (1.08%), and even this small fraction is enough to cause a rise in infections, which peaks at $t \sim 120$, beyond which time caseloads decline monotonically. While the non-poor are worse affected in this relapsed portion of the epidemic, when we consider the dynamics entirely, the poor are starkly worse off in terms of infection caseloads, with 68% of the poor having been infected as compared to 61% of non-poor. However, even as the poor population is worse affected, this scenario serves as a useful reminder that the non-poor cannot escape the impacts of infectious spread just because they isolate themselves effectively for a given time period, if indeed the infection is still active in the overall population at the end of this period.

This second peak can be explained using the mean-field analytical description of the model, where we find that once the intervention period is over, at time $t_{int} + t_{dur}$, and given the profile of recovered and susceptible population at that time, this results in an effective transmission rate $R_e(t_{int} + t_{dur}) > 1$ (Fig. 6). Given this $R_e(t)$, infections start rising once again, before reaching a peak and declining.

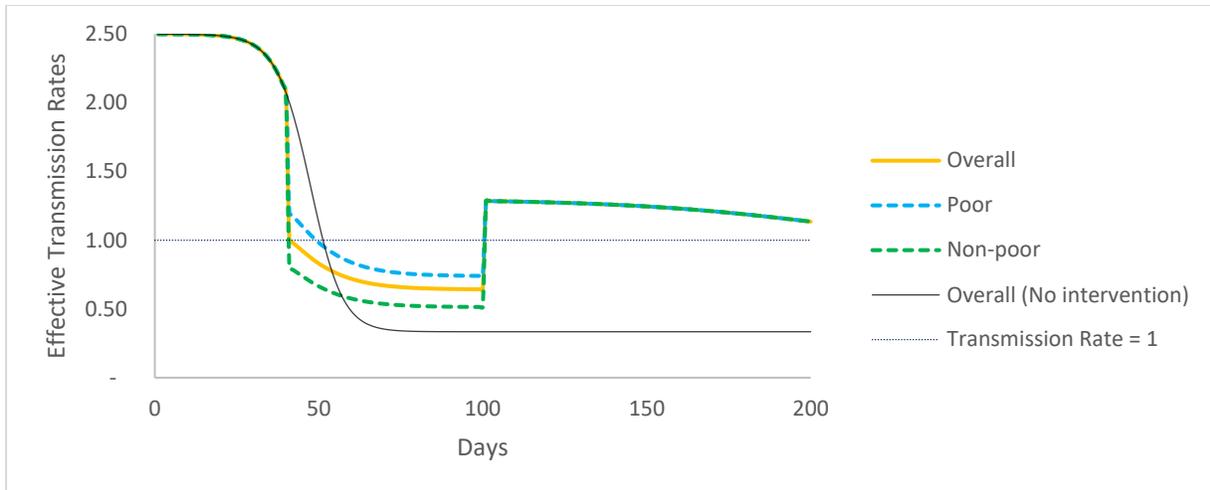

**Figure 6:** *Evolution of mean-field effective transmission rate under contact rate reduction scenario:* Until the intervention begins, effective transmission rate of all sections of the population follow the same trajectory, but then show differing paths with transmission rates for poor higher during the period of intervention. Post-intervention, again, the trajectories merge and effective transmission rate rises to above 1, meaning that infections are again taking hold in the population leading to the second, albeit smaller, peak in caseloads.

## 5. Sensitivity, robustness and interpretation:

Overall, the outcomes of our model suggest that the poor are worse off - in terms of both peak caseloads and cumulative infections - than non-poor under scenarios of contact rate reduction and transmission probability reduction implemented on the ER or BA network. Reducing either contact rate ($k$) or transmission probability ($p$), as we have done, are both mechanisms that effectively reduce $\beta$ (because $\beta = pk$), meaning that the daily transmission rate declines under all scenarios of intervention. The decline, though, is differential for the poor and the non-poor, causing the poor to have worse outcomes. This result also appears robust to changing the structure of the underlying network of disease transmission. We now turn to exploring differences in outcomes across various model parameters - simulating dynamics by varying total population, probability of being poor, recovery rate, and duration of intervention. All these simulations are done under the transmission probability ($p$) reduction scenario, because we have already seen that as long as the effect of the intervention is a reduction in $\beta$, the nature of outcomes remains consistent whether we reduce $p$ or $k$. Table 2 summarizes the results of simulations from 12 sets of parameter values for each network type.

| Parameter Type | Parameter Value | Differential Impact (DI) on ER network [95% CI] | Differential Impact (DI) on BA network [95% CI] |
| --- | --- | --- | --- |
| **Population** | 100,000 | 1.109 [1.098 - 1.121] | 1.049 [1.047 - 1.051] |
| | 10,000 | 1.101 [1.095 - 1.106] | 1.044 [1.042 - 1.046] |
| | 1,000 | 1.040 [1.024 - 1.056] | 1.036 [1.030 - 1.043] |
| **Proportion of poor** | 0.75 | 1.142 [1.121 - 1.163] | 1.059 [1.055 - 1.063] |
| | 0.50 | 1.101 [1.095 - 1.106] | 1.044 [1.042 - 1.046] |
| | 0.25 | 1.048 [1.042 - 1.055] | 1.028 [1.025 - 1.032] |
| **Recovery rate** | 0.05 | 1.017 [1.010 - 1.025] | 1.008 [1.007 - 1.009] |
| | 0.10 | 1.101 [1.095 - 1.106] | 1.044 [1.042 - 1.046] |
|  | 0.20 | 1.072 [1.046 - 1.097] | 1.099 [1.087 - 1.112] |
| Duration of Intervention | 30 (ER); 15 (BA) | 1.067 [1.063 - 1.072] | 1.028 [1.023 - 1.033] |
|  | 60 (ER); 30 (BA) | 1.101 [1.095 - 1.106] | 1.044 [1.042 - 1.046] |
|  | 120 (ER); 60 (BA) | 1.086 [1.076 - 1.096] | 1.045 [1.041 - 1.050] |

**Table 2**: *Simulation outcomes:* Outcomes on cumulative fractions of poor and non-poor infected under different parameter specifications of the model, with 95% Confidence Interval. Each parameter set is simulated 25 times, and the average outcomes are presented here. The Differential Impact is the ratio of the cumulative infected fraction of poor to non-poor.

Across all scenarios, the poor are worse affected than the non-poor. We also find that differential impacts are particularly sensitive to changes in certain parameters across both network implementations. Specifically, differential impacts are exacerbated - meaning that the impacts on the poor get progressively worse - with increases in population of the system and increases in proportion of poor population. It is also possible that differential impacts worsen with increasing duration of intervention, but this effect is not as obvious.

When we increase population, we keep the average size of an agent's neighbourhood ($q$) unchanged, meaning that each agent, on average, has the same number of contacts even as population increases an order of magnitude. By the time the intervention is initiated at $t_{int}$ on the ER graph, when $N = 1000$, already 59% of the cumulative poor and non-poor populations have been infected, when compared to 21% for $N = 10,000$ and 3% for $N = 100,000$. Therefore, when $N$ is smaller, there is a smaller fraction of the population post intervention left to be infected and for differentials to emerge, resulting in increasing differential impacts with increasing population. Before we can ascribe the rising differential impacts in this context to increasing population, it is important to unpack the effect of contact rates on this phenomenon. There are two cases with increasing average contact rate ($q$) and simultaneously increasing $N$ - one when $N$ increases by a factor of 10 (from 10,000 to 100,000) and $q$ by a factor of 5 (from 10 to 50) and the other when $N$ increases by a factor of 10 (from 1,000 to 10,000) and $q$ by a factor of 2 (from 50 to 100) – and we still find that differential impacts increase with increasing population in both cases. We simulate one final scenario where both $N$ and $q$ increase by an order of magnitude - $N$ increases by a factor of 10 (from 1,000 to 10,000) and $q$ by a factor of 10 (from 10 to 100) – and find that there is an increase in differential impact despite increasing $N$ and $q$ proportionally. This suggests that increasing population, irrespective of changes in contact rate, worsens the impact of epidemic spread on poor populations. It is anticipated that there will be over 40 megacities in the world by 2030 and most will be located in the developing world [18]. The worsening differential impact on the poor with increasing fraction of poor population therefore points to a significant public health challenge for these cities as they grow into the future.

Differential impact also shows a progressive increase with increasing fraction of poor. Given a random network, the likelihood of contact between two poor agents, on average, is $f_p^2(1 - f_S)$, and the effective transmission rate between two poor agents, after the intervention occurs, is $f_p^2(1 - f_S)\beta_{p-p}$. As the fraction of poor, $f_p$, increases (from 0.25 to 0.5 to 0.75 in our simulations), the effective transmission rate between poor agents, $f_p^2(1 - f_S)\beta_{p-p}$ increases as well. Due to this, the infection rates of the poor increase with increasing poor population, resulting in increasing differential impact between poor and non-poor (as $\beta_{p-p} > \beta_{p-np}, \beta_{np-p}, \beta_{np-np}$).

Increasing differential impact with increase in duration of intervention could occur because, as just discussed, transmission rates between poor agents is higher than transmission rates between all other combinations of agent types for the duration of interventions. In longer interventions, therefore, progressively greater fractions of poor are infected compared to the non-poor. Therefore, the poor are likely to contend with both the aggravated negative health and economic consequences of an epidemic at the same time.

Finally, it is apparent that increasing recovery rates indicate decreasing $R_0$, meaning that the average number of infections per infective declines. Therefore, we would expect that as $\gamma$ increases, the caseload of infected decreases. This is the case in both the ER and BA graphs where we find that as $\gamma$ increases from 0.05 to 0.1 to 0.2, the overall fraction of infected decreases - from 0.88 to 0.80 to 0.45 for the ER network, and from 0.99 to 0.93 to 0.76 for the BA network. The higher corresponding infected fractions for the BA network (for a given value of $\gamma$) reflect the disproportionately important role of 'super-spreader' or highly connected nodes in the network of spread. While differential impact increases with increasing $\gamma$ in the BA network, it increases from $\gamma = 0.05$ to $\gamma = 0.1$, but then declines for $\gamma = 0.2$ in the ER network. A likely reason for this discrepancy is the significantly lower fraction of overall infections in the ER network (0.45), compared to 0.76 in the BA network for $\gamma = 0.2$, which potentially mitigates against large differentials emerging in the ER network. We find evidence for this when we simulate model dynamics for intermediate values of $\gamma$ ($= 0.15, 0.18$) and find that while differential impact is highest for the ER network at 1.123 for $\gamma = 0.15$, it declines to 1.079 at $\gamma = 0.18$, when the average fraction of total infected is ~45%. This suggests that differential impact does not monotonically increase with increasing $\gamma$, but that it peaks and then declines when $\gamma$ corresponds to a certain threshold of total fraction of population infected.

Although the poor are worse off in all the scenarios we have looked at, we cannot afford to see the problem as confined in relevance to the poor population alone. Fig. 6a shows how the unequal dynamics of the epidemic makes it possible for the disease to linger amongst the poor and then suddenly be transmitted to the large susceptible population of the non-poor, which were not infected in the first round of the epidemy. This implies that policies addressing differential impacts are not just relevant for the poor, but critical for society as a whole.

## 5.Discussion:

Emerging evidence on the Covid-19 pandemic suggests that socio-economic deprivation and overcrowding are resulting in Black and Asian communities in the UK, slum dwellers in India, and African-American and Latino individuals in the US, being disproportionately affected [19, 20, 21, 22, 23, 24]. There is concern that the causal mechanism between poverty and susceptibility to infection includes not only previous medical history, but also the particular social and economic contexts of overcrowded accommodations, economic compulsions (making it difficult to stay at home), increased stress levels, poverty induced comorbidities, and reduced access to health care [19, 20]. Long term resilience to epidemics therefore requires both pharmaceutical interventions (such as medicines and vaccines to particular episodes of epidemy) and broader socioeconomic action on foundational concerns of housing, access to health care, and income fragility.

Our work finds that increasing populations and increasing fractions of poor population are likely to exacerbate epidemy outcomes for the poor in urban contexts, and that therefore we need to better equip public health systems in such contexts to deal with epidemic outbreaks. Specifically, in the context of the developing world, where hospital infrastructure is severely inadequate, it is argued that information, communication, and primary public health infrastructure could meaningfully be leveraged to contain epidemics [21, 23]. There is a need for systematic focus on studying the impacts of infectious outbreaks on economically vulnerable communities so as to more rigorously identify effects of community, health systems, and policy [21]. Often in low- and medium-income countries, the strengths of health systems lie in networks of ground-level community level health workers, and it is this network of primary care providers that can be leveraged for more targeted and meaningful communication, infection monitoring, and with appropriate training, testing and evaluation [23]. Aspects of such a strategy were visible in the containment of spread of Covid-19 in Mumbai's Dharavi slum [24]. Finally, sustained improvements in the provision of basic drinking water and toilet facilities to the entire urban population will enable creation of healthier urban spaces, which will also aid in combating spread of infections.

## 6. Conclusion:

We simulate the emergence of differential outcomes for poor and non-poor populations in a city using a simple network model of epidemic spread. We study the fractions of poor and non-poor infected in the course of an epidemic under different kinds of intervention, and across a range of parameter values. Under the non-intervention scenario, we find that peak infection caseload is maximised, but that there are no differences in infection levels between the poor and non-poor populations. Once we have an intervention, it serves to reduce peak caseload in the system, but we find the poor are now consistently worse off than the non-poor under all intervention scenarios, and irrespective of the underlying network structure. We also find that increasing the city population, fraction of the urban poor, or duration of intervention leads to a progressive worsening of outcomes for the poor vis-à-vis the non-poor. Policies focused on reducing differential impacts of epidemic spread benefit not just the poor, but improve health outcomes for the entire population. Such policies include the development of a rigorous understanding of links between poverty and epidemics, as well as leveraging extant primary care infrastructure in developing countries.

**Author Contributions:** AS and HJJ conceived the research, chose the methodology, wrote the analytical description, and reviewed and edited the manuscript. AS programmed the simulation and wrote the draft manuscript.

**Competing Interests:** The authors declare that they have no competing interests.

**Funding:** The authors received no funding for this work.